\documentclass[showpacs,preprintnumbers,amsmath,amssymb,prb,twocolumn,superscriptaddress]{revtex4}

\usepackage{graphicx}
\usepackage{dcolumn}
\usepackage{bm}

\begin{document}

\title{Spin and Charge Josephson effects between non-uniform
superconductors with coexisting helimagnetic order}

\author{Ilya Eremin}
\affiliation{Max-Planck
Institut f\"ur Physik komplexer Systeme, N\"othnitzerstr 38,
D-01187 Dresden, Germany}
\affiliation{Institut f\"ur Mathematische/Theoretische Physik,
Technische Universit\"at Carolo-Wilhelmina zu
Braunschweig, D-38106 Braunschweig, Germany}
\author{Flavio S. Nogueira}
\affiliation{Institut f\"ur Theoretische Physik,
Freie Universit\"at Berlin, Arnimallee 14, D-14195 Berlin, Germany}
\author{Ren\'e-Jean Tarento}
\affiliation{Laboratoire de Physique des Solides, UMR 8502 - Universit\'e
Paris-Sud, B\^at. 510, F-91405 Orsay Cedex, France}

\date{Received \today}

\begin{abstract}
 We consider the spin and charge Josephson current between two non-uniform
Fulde-Ferrel-Larkin-Ovchinnikov superconductors with helimagnetic order. We demonstrate
that the presence of the
helimagnetic phase generates a spin Josephson effect and leads to additional
contributions  to both single-particle and Josephson charge current. It is
shown that for such systems the AC effect differs more radically from the DC
effect than in the case of a BCS superconductor with
helimagnetic order considered earlier in
the literature [M. L. Kuli\'c and I. M. Kuli\'c, Phys. Rev. B {\bf 63},
104503 (2001)] where a spin Josephson current has also been found.
In our system the most interesting effect
occurs in the presence of an external magnetic field and in
absence of voltage, where we show that the charge Josephson current can be tuned to zero
while the spin Josephson current is non-vanishing. This provides a well controlled
mechanism to generate a spin supercurrent in absence of charge currents.
\end{abstract}


\maketitle

\section{Introduction}

Collective spin and charge
transport phenomena in ordered many-particle systems are
of great importance
in modern condensed matter physics.
Among them is the dissipationless, Cooper-pair driven, transport
in superconductors and in the superfluid $^3$He. One
remarkable consequence of the supercurrent phenomenon is the
Josephson effect,\cite{review1} which predicts that a
a superflow exists between two superfluid systems
(charged or not) separated by a
weak link, and that its value is proportional to the
sine of the difference between the phases of the complex order parameter
across the link.

Recently, due to the growing interest in spintronics devices,
there were a number of works exploring the possibility for
dissipationless spin current.\cite{nagaosa,kopietz,konig,flavio,review2} In
singlet superconductors it cannot occur because the total spin
of the Cooper-pair is zero. However, in unconventional triplet
superconductors this may not be the case.
Moreover, the B-phase of superfluid
$^3$He exhibits both
mass and spin-1 supercurrents. The latter was probed in
an experiment where two $^3$He-B superfluids were
in contact through a weak-link.\cite{bunkov} This
led to the observation of a spin Josephson effect, thus
establishing the existence of spin supercurrents in the
B-phase of superfluid  $^3$He.
More recently it was pointed out that a spin
Josephson effect between two triplet
ferromagnetic superconductors may occur.\cite{sudbo}

In the above mentioned scenarios of phase coherence in systems
with fermionic pairing the order parameter is uniform.
Superconductivity with non-uniform order parameter occurs, for example,
in the presence of an exchange field.
This class of superconductors is well described by the
so called Fulde-Ferrel-Larkin-Ovchinnikov (FFLO) state.\cite{fulde,larkin}
Experimentally, it should occur in
extremely high-field superconductors, which are obviously of high
practical use. Recently, strong evidence has been found
that a FFLO state might be realized in the quasi-two-dimensional
heavy-fermion superconductor CeCoIn$_5$ \cite{exp1,exp2} for a
magnetic field applied along the $ab$-plane. In this respect, the
coexistence of a helimagnetic phase induced by the in-plane
magnetic field, or as an intrinsic order parameter,
may result in various interesting transport
phenomena similar to some of the $^3$He features,
though the system considered is in a singlet
state.

In this paper we analyse the spin and charge tunnelling processes between
two FFLO-like helimagnetic superconductors and find that the
spin-flip processes associated to the helimagnetic phase result in
spin and charge tunnelling of the Josephson type, i.e., the phase
differences of the superconducting and helimagnetic orders are
involved in the tunnel process. Previously  a similar
analysis was undertaken for helimagnetic superconductors with a
uniform superconducting order parameter.\cite{Kulic} In the case
of vanishing voltage our results reduce essentially to the ones
obtained in Ref. \onlinecite{Kulic}. However, at nonzero voltages the
corresponding AC Josephson effect in non-uniform helimagnetic
superconductors changes the physics more drastically than in the
case of BCS superconductors. We will also study the effect of a nonzero
magnetic field in the Josephson currents at zero voltage. It will be
shown that the magnetic field can be used to tune the charge Josephson
current to zero, while still having a nonzero spin Josephson current. In
this way we provide a mechanism by which a spin supercurrent exists in
the absence of charge currents.

The plan of the paper is as follows. In Sect. II we briefly discuss the
Josephson effect between two FFLO superconductors from the Ginzburg-Landau
theory for the FFLO state. Sect. III introduces our model for a non-uniform
helimagnetic superconductor. There we derive the Green functions of the theory
using mean-field approach. In Sect. IV the single-particle and
Josephson charge and spin currents are derived using linear response.
The effect of an external magnetic field is considered in Sect. V. Our
conclusions are presented in Sect. VI.

\section{Josephson effect between two
non-uniform superconductors}

In order to gain some insight on the physics of the Josephson
effect between non-uniform superconductors,
let us consider first a FFLO superconductor, where the exchange field is
uniform. For this situation a Ginzburg-Landau (GL) free energy has been
derived in Ref. \onlinecite{Buzdin}. We can use this result to obtain the charge
supercurrent in the absence of magnetic field:\cite{Buzdin}
\begin{eqnarray}
\label{GLcur}
{\bf j}&=&-i2e[\beta+(\mu-2\eta)|\psi|^2](\psi^*\nabla\psi-\psi\nabla\psi^*)
\nonumber\\
&-&4ie\delta(\nabla\psi^*\nabla^2\psi-\nabla\psi\nabla^2\psi^*),
\end{eqnarray}
where $\beta$, $\mu$, $\eta$, and $\delta$ are phenomenological parameters. When two such
FFLO superconductors are connected through a tunnel junction, we can consider the
charge flow from the left to the right subsystems with appropriate boundary
conditions at the tunnel junction. The procedure is similar to the case
of ordinary uniform superconductors, except that the above current must be used
instead. Let us  consider the $x$-component of the
current ${\bf j}_L$ coming from the left side of
the junction, which is given by the expression
(\ref{GLcur}) along the
$x$-direction, with $\psi$ replaced by $\psi_L$. At lowest
order we have the boundary conditions $\partial_x\psi_L=\lambda\psi_R$ and
$\partial_x^2\psi_L=\lambda\partial_x\psi_R$ at the tunnel junction, where
$\psi_R$ is the order parameter of the right subsystem and $\lambda$ is
a parameter depending on the details of the junction. By assuming an order
parameter of the Fulde-Ferrel type, we can approximately write
$\psi_L({\bf r})=\rho_0 e^{i(\theta_L+{\bf q}\cdot{\bf r})}$
and  $\psi_R({\bf r})=\rho_0 e^{i(\theta_R+{\bf q}\cdot{\bf r})}$,
with $\rho_0={\rm const}$. Note that
we are assuming that both sides are made with the same material, so that the
amplitude $\rho_0$ is the same on either side. The current flowing through the junction is
then
\begin{equation}
\label{jL}
j_{Lx}=4e\lambda\rho_0^2[2\delta q_x^2+\beta+(\mu-2\eta)\rho_0^2]\sin\Delta\theta,
\end{equation}
where $\Delta\theta\equiv\theta_R-\theta_L$.
Note that the amplitude of the Josephson current depends on the FFLO
characteristic momentum.
The presence of a Josephson effect between
two FFLO superconductors is in contrast with the situation of a
tunnel junction
between a FFLO superconductor and a superconductor having a uniform order
parameter. In such a case, it can be shown that the Josephson effect is
suppressed, since the uniform state is not able to balance the spatial
oscillations from the FFLO state.\cite{Agterberg}

\section{Non-uniform
helimagnetic superconductors}

\subsection{Helimagnetic superconductors}

The study of helimagnetic superconductors has a long story, mainly
associated with heavy fermion materials. Particularly interesting is the
following model introduced long time ago,\cite{Varma} whose free energy is
given by
\begin{eqnarray}
\label{crypto}
{\cal F}&=&|(\nabla-i2e{\bf A})\psi|^2+a|\psi|^2+\frac{b}{2}|\psi|^4
\nonumber\\
&+&\frac{1}{2}(\nabla{\bf M})^2+\frac{r}{2}{\bf M}^2+
\frac{u}{8}({\bf M}^2)^2\nonumber\\
&+&\frac{1}{8\pi}(\nabla\times{\bf A}-4\pi{\bf M})^2
\end{eqnarray}
where ${\bf M}$ is the macroscopic magnetisation. The above free
energy admits a mean-field solution with a helical magnetically
ordered state and a uniform superconducting order parameter. A non-uniform
order parameter of the Fulde-Ferrel type,
$\psi({\bf r})=\psi_0 e^{i{\bf q}\cdot{\bf r}}$, does not work in this
case, since the wave-vector ${\bf q}$ can be gauged away through a
gauge transformation ${\bf A}\to{\bf A}+{\bf q}/2e$. Thus, for the
above model a possible  non-uniformity of the superconducting order
parameter does not contain any additional physics with respect to the
uniform case, at least not at the macroscopic level.

A mean-field microscopic model having a uniform superconducting
order parameter
and helimagnetic order would have, in an easy-plane configuration, the
Hamiltonian
\begin{eqnarray}
\label{H-hel}
H&=&\frac{1}{2m}\sum_\sigma\nabla c_\sigma^\dagger({\bf r})\cdot
\nabla c_\sigma({\bf r})-\mu\sum_\sigma c_\sigma^\dagger({\bf r}) c_\sigma({\bf r})
\nonumber\\
&+&\Delta_0c_\uparrow^\dagger({\bf r})c_\downarrow^\dagger({\bf r})
+h_{\bf q}e^{i{\bf q}\cdot{\bf r}}
c^\dagger_\uparrow({\bf r})c_\downarrow({\bf r})+{\rm h.c.}
\end{eqnarray}
The transformation $c_\sigma\to e^{i\sigma{\bf q}\cdot{\bf r}/2}c_\sigma$
produces a spin current response, since the Hamiltonian becomes
\begin{eqnarray}
\label{H-hel-1}
\lefteqn{H = \frac{1}{2m}\sum_\sigma\nabla c_\sigma^\dagger({\bf r})\cdot
\nabla c_\sigma({\bf r})-\left(\mu-\frac{q^2}{8m}\right) \times} &&
\nonumber\\
&& \sum_\sigma c_\sigma^\dagger({\bf r}) c_\sigma({\bf r})+
\Delta_0c_\uparrow^\dagger({\bf r})c_\downarrow^\dagger({\bf r})
+h_{\bf q}
c^\dagger_\uparrow({\bf r})c_\downarrow({\bf r})+{\rm h.c.}\nonumber\\
&& - {\bf q}\cdot({\bf j}_\uparrow-{\bf j}_\downarrow),
\end{eqnarray}
where
\begin{equation}
{\bf j}_\sigma=\frac{i}{4m}[c_\sigma^\dagger\nabla c_\sigma
-(\nabla c_\sigma^\dagger)c_\sigma],
\end{equation}
is the current for the spin $\sigma$ fermion. Thus, although
in the above microscopic
model there are no charge currents in the absence of electromagnetic
coupling (the momentum of the Cooper pairs is zero), there is a
spin current. However, if in the Hamiltonian (\ref{H-hel})
$\Delta_0$ is replaced by a Fulde-Ferrel mean-field order parameter
$\Delta_{\bf p} e^{i{\bf p}\cdot{\bf r}}$ and the transformation
$c_\sigma\to e^{i({\bf p}+\sigma{\bf q}\cdot{\bf r}/2)}c_\sigma$ is done,
we obtain a charge current response in addition to the spin current one,
i.e.,
\begin{eqnarray}
\label{H-hel-2}
\lefteqn{H = \frac{1}{2m}\sum_\sigma\nabla c_\sigma^\dagger({\bf r})\cdot
\nabla c_\sigma({\bf r})-\sum_\sigma\left[\mu-\frac{({\bf p}+\sigma{\bf q})^2 }{8m}\right]
\times}  &&
\nonumber\\
&& c_\sigma^\dagger({\bf r}) c_\sigma({\bf r}) + \Delta_{\bf p}c_\uparrow^\dagger({\bf r})c_\downarrow^\dagger({\bf r})
+h_{\bf q}
c^\dagger_\uparrow({\bf r})c_\downarrow({\bf r})+{\rm h.c.}\nonumber\\
&& - {\bf q}\cdot({\bf j}_\uparrow-{\bf j}_\downarrow)
-{\bf p}\cdot({\bf j}_\uparrow+{\bf j}_\downarrow).
\end{eqnarray}
From the above equation we see that there is an excess kinetic
energy of amount $({\bf p}+{\bf q})^2/(8m)$ and $({\bf p}-{\bf
q})^2/(8m)$ for the spin up and down electrons, respectively.
Setting ${\bf p}={\bf q}$ has the effect of producing a (charge)
current response only for the up spin electrons while adding no
extra kinetic energy to the down spin electrons. The situation in
such a state is the one similar to injecting fully polarized
electrons in a sample. An important additional property of the ${\bf
p}={\bf q}$ state is that the magnetic order parameter $\langle
c_{{\bf k+q/2}\uparrow}^\dagger c_{{\bf k-q/2}\downarrow} \rangle$
can be transformed in the superconducting one $\langle c_{{\bf
k+q/2}\uparrow}^\dagger c_{{\bf -k+q/2}\downarrow}^\dagger \rangle$
through a particle-hole transformation in the down spin channel,
i.e., $c_{{\bf k-q/2}\downarrow}\to c_{{\bf
-k+q/2}\downarrow}^\dagger$. In other words, the corresponding order
parameters can be rotated into one another.  The magnetic and
superconducting order parameters with a same helical pattern are
more coherent: a non-uniform Cooper pair breaking is likely to imply
a decay into the helimagnetic state. Due to these interesting
properties, we will consider from now on a mean-field microscopic
model where both superconducting and magnetic order parameters have
the same helical pattern.

Finally, we would like to stress here that there is no contradiction between
Eqs. (\ref{crypto}) and (\ref{H-hel-2}). As a matter of fact,
Eq. (\ref{crypto}) is not suitable to describe a superconductor
with a non-uniform superconducting order parameter.
Instead, another form of the free energy
has to be used,\cite{Buzdin} since in this case higher order derivatives have to be
taken into account.

\subsection{Green functions for non-uniform
helimagnetic superconductors}

Following th discussion of the previous Subsection,
let us consider a FFLO-like superconductor with
a helimagnetic molecular field. The mean-field Hamiltonian is given
in momentum space by
\begin{eqnarray}
\label{MFH}
H_{\rm MF}&=&\sum_{{\bf k} \sigma } \varepsilon_{\bf k}
c^{\dagger}_{{\bf k}\sigma} c_{{\bf k}\sigma}
+\sum_{\bf k} \Delta^*_{\bf q} c_{{\bf -k+q/2}\downarrow}c_{{\bf k+q/2}
\uparrow}
\nonumber\\
&+&  \sum_{\bf k} h_{\bf q} c_{{\bf k+q/2}\uparrow}^\dagger c_{{\bf k-q/2}\downarrow}
+ {\rm h.c.} \quad,
\end{eqnarray}
where $\epsilon_{\bf k}$ is the quadratic dispersion of the free
electrons. We are assuming that the oscillation of the
superconducting condensate is characterised by a single wave
vector {\bf q}, i.e., $\psi ({\bf r}) \propto  e^{\bf i{\bf
q}\cdot{\bf r}}$.\cite{fulde} $h_{\bf q}$ is a
complex mean-field variable describing the helimagnetic phase
characterised by the electron-hole singlet pairing $\langle c_{{\bf
k+q/2}\uparrow}^\dagger c_{{\bf k-q/2}\downarrow} \rangle$. As already discussed,
{\it in our model both the superconducting
and helimagnetic order parameters are modulated
by the same wave-vector} ${\bf q}$. This is important, since for ${\bf q}=0$
no coexistence between magnetic and superconducting order will be possible
within our model.

Generally, helimagnetism can be induced by the external
inhomogeneous magnetic field applied along the $x$-direction or
arising from internal (spiral) magnetic order. In the absence of superconductivity, our
theory reduces to the one considered in Ref. \onlinecite{konig},
which corresponds to a magnetic analog of the FFLO state. There the
existence of persistent spin currents was demonstrated.

From the mean-field Hamiltonian we see that in the present case
not only the gauge symmetry is spontaneously broken, leading to a
lack of particle number conservation, but also the spin
conservation symmetry is broken due to the
helimagnetic phase. Both averages are complex and have therefore
an amplitude and a phase, i.e., $\Delta_{\bf q} = |\Delta_{\bf
q}|e^{-i\theta}$ and $h_{\bf q} = |h_{\bf
q}|e^{-i\varphi}$. In a bulk system both
phases can be gauged away through a global gauge transformation.
This is of course not the case when we consider the tunnelling
processes between two superconductors and, as we will show later,
the phase of the helimagnetic order parameter will play
an important role.

The mean-field Hamiltonian (\ref{MFH}) can be
conveniently rewritten in matrix form as
$H_{\rm MF}=(1/2)\sum_{\bf k}\eta_{\bf k}^\dagger M_{\bf k}\eta_{\bf k}$,
where $\eta_{\bf k}^\dagger=[c_{{\bf k+q/2}\, \uparrow}^\dagger~~
c_{{\bf k-q/2}\,\downarrow}^\dagger~~c_{{\bf -k+q/2}\,\downarrow}~~
c_{{\bf -k-q/2}\,\uparrow}]$
and
\begin{equation}
M_{\bf k}=\left[
\begin{array}{cccc}
\varepsilon_{\bf k+q/2} & |h_{\bf q}|e^{-i\varphi} & |\Delta_{\bf q}|e^{-i\theta}
& 0\\
\noalign{\medskip}
|h_{\bf q}|e^{i\varphi} & \varepsilon_{\bf k-q/2} & 0 & - |\Delta_{\bf q}|
e^{-i\theta}\\
\noalign{\medskip}
|\Delta_{\bf q}|e^{i\theta} & 0 & -\varepsilon_{\bf -k+q/2} &
-|h_{\bf q}|e^{-i\varphi}\\
\noalign{\medskip}
0 & -|\Delta_{\bf q}|e^{i\theta} & -|h_{\bf q}|e^{-i\varphi} &
-\varepsilon_{\bf - k-q/2}
\end{array}
\right].
\end{equation}
The matrix $M_{\bf k}$ can be easily diagonalized, which leads to the
following energy spectrum
\begin{equation}
\label{spectrum}
E^{\alpha,\beta}_{{\bf k}}=\alpha\sqrt{\varepsilon_{{\bf k}\,s}^2+
|\Delta_{\bf q}|^2}
+\beta \sqrt{\varepsilon_{{\bf k}\,a}^2+|h_{\bf q}|^2},
\end{equation}
and $\alpha, \beta=\pm 1$. Here
we introduce $\varepsilon_{{\bf k}\,s}= (
\varepsilon_{{\bf k+q/2}} +\varepsilon_{{\bf k-q/2}})/2$ and
$\varepsilon_{{\bf k}\,a}= (
\varepsilon_{{\bf k+q/2}} - \varepsilon_{{\bf k-q/2}})/2$
similarly to Refs. \onlinecite{shimahara} and
\onlinecite{takada}. Since we assume the quadratic
dispersion for the free electrons, we have
$\varepsilon_{{\bf k}\,s}=\varepsilon_{\bf k} + q^2/8m$
and $\varepsilon_{{\bf k}\,a}=\frac{v_{F}q}{2} \cos x$ and $x$ is the angle
between {\bf k} and {\bf q}.

The matrix Green function is obtained by inverting the
matrix $-i\omega I+M_{\bf k}$, where $I$ is the identity matrix. The
independent elements of the matrix Green function are
\begin{widetext}
\begin{equation}
G^{\uparrow,\uparrow}_1(i\omega_n,{\bf k})\equiv\langle c_{{\bf
k+q/2}\uparrow}^\dagger(i\omega) c_{{\bf
k+q/2}\uparrow}(i\omega)\rangle =  \frac{({u^{++}_{\bf
k}})^2}{i\omega_n-E_{1{\bf k}}} + \frac{({u^{--}_{\bf
k}})^2}{i\omega_n+E_{1{\bf k}}} + \frac{({u^{+-}_{\bf
k}})^2 }{i\omega_n-E_{2{\bf k}}} +
\frac{({u^{-+}_{\bf
k}})^2 }{i\omega_n+E_{2{\bf k}}} \quad,
\end{equation}
\begin{equation}
G^{\uparrow,\downarrow}_2(i\omega_n,{\bf k})\equiv\langle c_{{\bf
k+q/2}\uparrow}^\dagger(i\omega) c_{{\bf
k-q/2}\downarrow}(i\omega)\rangle  =  - \frac{h_{\bf
q}e^{-i\varphi}\left(E_{1{\bf k}}E_{2{\bf
k}}+2(i\omega_n)\varepsilon_{{\bf k}s}+ (i\omega_n)^2\right)
}{(i\omega_n-E_{1{\bf k}})(i\omega_n-E_{2{\bf k}})
(i\omega_n+E_{1{\bf k}})(i\omega_n+E_{2{\bf k}})} \quad,
\end{equation}
\begin{equation}
F^{\uparrow,\downarrow}_1(i\omega_n,{\bf k})\equiv\langle c_{{\bf
k+q/2}\uparrow}^\dagger(i\omega) c_{{\bf
-k+q/2}\downarrow}^\dagger(-i\omega)\rangle  =  -
\frac{\Delta_{\bf q} e^{-i\theta}\left( (i\omega_n)^2-E_{1{\bf
k}}E_{2{\bf k}}-2(i\omega_n)\varepsilon_{{\bf
k}a}\right)}{(i\omega_n-E_{1{\bf k}})(i\omega_n-E_{2{\bf k}})
(i\omega_n+E_{1{\bf k}})(i\omega_n+E_{2{\bf k}})} \quad,
\end{equation}
\begin{equation}
F^{\uparrow, \uparrow}_2(i\omega_n,{\bf k})\equiv\langle c_{{\bf
k+q/2}\uparrow}^\dagger(i\omega) c_{{\bf
-k-q/2}\uparrow}^\dagger(-i\omega)\rangle  =  \frac{2h_{\bf
q}e^{-i\varphi} \Delta_{\bf q}e^{-i\theta}(i
\omega_n)}{(i\omega_n-E_{1{\bf k}})(i\omega_n-E_{2{\bf k}})
(i\omega_n+E_{1{\bf k}})(i\omega_n+E_{2{\bf k}})} \quad,
\label{gf}
\end{equation}
where we have set $E_{1{\bf k}}\equiv E^{+,+}_{{\bf k}}$ and
$E_{2{\bf k}}\equiv E^{+,-}_{{\bf k}}$ and the generalized Bogolyubov coefficients
are
\begin{equation}
u_{\bf k}^{\alpha\beta}=\frac{1}{2}\left[1 +(2\delta_{\alpha\beta}-1)
\frac{4\varepsilon_{{\bf k}a}
\varepsilon_{{\bf k}s}}{E_{1{\bf k}}^2-E_{2{\bf k}}^2} + \frac{2\alpha
\varepsilon_{{\bf k}s}}{E_{1{\bf k}}+E_{2{\bf k}}}+ \frac{2\beta
\varepsilon_{{\bf k}a}}{E_{1{\bf k}}-E_{2{\bf k}}}
\right]^{1/2} \quad,
\end{equation}
where $\alpha,\beta=\pm 1$.

\section{Josephson effect between two non-uniform
helimagnetic superconductors}

We will study first the charge single-particle and
Cooper-pair Josephson tunnelling processes between two FFLO
superconductors. We use the standard tunnelling
Hamiltonian \cite{josephson} in the form
\begin{equation}
H_T = \sum_{{\bf k\, p }\, \sigma} T_{\bf k,p}c_{{\bf k}\sigma}^{\dagger}
c_{{\bf p}\sigma} +{\rm h.c.}  \quad,
\end{equation}
where {\bf k} and {\bf p} label single electron momentum
eigenstates in the left and right subsystems,
respectively. The charge current is given by $I^{\rm charge}= -e \langle
\dot{N}_L(t) \rangle$, where $N_L= \sum_{{\bf k}, \sigma} c_{{\bf
k}\sigma}^{\dagger} c_{{\bf k}\sigma}$ and the spin current is
given by $I^{\rm spin}= -\mu_B \langle \dot{S}_z(t) \rangle$, where
$S_z= \sum_{{\bf k}, \sigma} \sigma c_{{\bf k}\sigma}^{\dagger}
c_{{\bf k}\sigma}$ and $\mu_B$ is the Bohr magneton. In the linear
response regime the charge and spin currents are given by
$I_s^{\rm charge}= 2e {\rm Im} [X_{\rm charge}(eV+i\delta)]$ and $I_s^{\rm spin}=
2\mu_{B} {\rm Im} [X_{\rm spin}(eV+i\delta)]$, where $\delta \to 0^+$. In
terms of the Matsubara formalism at lowest order one gets:
%
%
\begin{eqnarray}
\lefteqn{X_{\rm charge,spin}(i\omega)=-\frac{1}{\beta}\sum_{\omega_n}\sum_{{\bf
k},{\bf p}} \left[|T_{\bf k+q/2, p+q/2}|^2 G_1^{\uparrow,
\uparrow}({\bf k}, i\omega_n)G_1^{\uparrow, \uparrow}({\bf p},
i\omega_n-i\omega_m) \right. }
&& \nonumber\\
&& \left. \pm |T_{\bf k-q/2, p-q/2}|^2 G_1^{\downarrow,
\downarrow}({\bf k}, i\omega_n)G_1^{\downarrow, \downarrow}({\bf
p}, i\omega_n-i\omega_m) + T_{\bf k+q/2, p+q/2}T^*_{\bf k-q/2,
p-q/2} \times \right. \nonumber\\
&& \left. G_2^{\uparrow, \downarrow}({\bf k},
i\omega_n)G_2^{\downarrow, \uparrow}({\bf p}, i\omega_n-i\omega_m)
\pm  T_{\bf k-q/2, p-q/2}T^*_{\bf k+q/2, p+q/2}G_2^{\downarrow,
\uparrow}({\bf k}, i\omega_n)G_2^{\uparrow, \downarrow}({\bf p},
i\omega_n-i\omega_m) \right] \quad,
\end{eqnarray}
%
%
where $+$, $-$ refer to the charge and spin current, respectively.
Besides the usual contribution to the
single particle charge current involving the product of Green
functions $G_{\uparrow \uparrow}$ and $G_{\downarrow \downarrow}$
from the left and right sides of the junction, there are
extra contributions involving the Green's functions $G_{\uparrow
\downarrow}$ and $G_{\downarrow \uparrow}$ which give a
term proportional to $e^{i(\varphi_L-\varphi_R)}$. In particular,
after straightforward calculations we get for the single-particle charge
current
\begin{equation}
I_s^{\rm charge}(eV)=I_0(eV)+I_1(eV)\cos\Delta\varphi \quad,
\end{equation}
where $I_0(eV)$ is the single particle current
%
\begin{eqnarray}
I_0(eV) & = &  2 \pi e|T|^2 \sum_{{\bf k},{\bf p},i\neq j} \left(
1+\frac{\varepsilon_{{\bf k}a}\varepsilon_{{\bf
p}a}}{\sqrt{\varepsilon_{{\bf k}a}^2+|h_{\bf
q}|^2}\sqrt{\varepsilon_{{\bf p}a} +|h_{\bf q}|^2}}\right) \left[
\delta(eV-E_{i{\bf k}}-E_{i{\bf p}})-
\delta(eV+E_{i{\bf k}}+E_{i{\bf p}})\right] \nonumber \\
&& + \left( 1-\frac{\varepsilon_{{\bf k}a}\varepsilon_{{\bf
p}a}}{\sqrt{\varepsilon^2_{{\bf k}a}+|h_{\bf
q}|^2}\sqrt{\varepsilon_{{\bf p}a}+|h_{\bf q}|^2}}\right) \left[
\delta(eV-E_{i{\bf k}}-E_{j{\bf p}}) -\delta(eV+E_{i{\bf
k}}+E_{j{\bf p}})\right]
\end{eqnarray}
%
and for $I_1^{\rm charge}(eV)$ we find
\begin{eqnarray}
I_1(eV)  =  2 \pi e|T|^2 \sum_{{\bf k},{\bf p},i\neq j} \left(
\frac{|h_{\bf q}|^2}{\sqrt{\varepsilon_{{\bf k}a}^2+|h_{\bf
q}|^2}\sqrt{\varepsilon_{{\bf p}a}+|h_{\bf q}|^2}}\right) &&
\left[ \delta(eV-E_{i{\bf k}}-E_{i{\bf p}})-
\delta(eV+E_{i{\bf k}}+E_{i{\bf p}}) \right.-\nonumber \\
&& \left.\delta(eV-E_{i{\bf k}}-E_{j{\bf p}})- \delta(eV+E_{i{\bf
k}} +E_{j{\bf p}})\right]
\end{eqnarray}
%
%
 while
for the single-particle spin current we obtain
\begin{equation}
I_s^{\rm spin}(eV)=\tilde I_0(eV)\sin\Delta\varphi.
\end{equation}
with
\begin{eqnarray}
\tilde I_0(eV)   & = &  2 \mu_B |T|^2 \sum_{{\bf k},{\bf p},i\neq
j} \left( \frac{|h_{\bf q}|^2}{\sqrt{\varepsilon^2_{{\bf
k}a}+|h_{\bf q}|^2}\sqrt{\varepsilon^2_{{\bf p}a}+|h_{\bf
q}|^2}}\right) \times
\nonumber \\
&& \left[ \frac{1}{eV+E_{i{\bf k}}+E_{i{\bf p}}} -
\frac{1}{eV-E_{i{\bf k}}-E_{i{\bf p}}} - \left(
\frac{1}{eV+E_{i{\bf k}}+E_{j{\bf p}}} - \frac{1}{eV-E_{i{\bf
k}}-E_{j{\bf p}}} \right) \right]
\end{eqnarray}
%

The first difference in the single-particle charge current occurs
due to the FFLO state itself which would be present also in the
case of the uniform exchange field. Most interestingly, the
presence of the helimagnetic phase and the corresponding breaking
of the SU(2) symmetry induces an additional term in the
single-particle charge current proportional to $\cos \Delta
\varphi$ and generated the corresponding term in the spin current
proportional to $\sin \Delta \varphi$. The form of the
single-particle spin current resembles the one in the so-called
``spin Josephson effect'' in ferromagnetic/ferromagnetic junctions
\cite{flavio} which, strictly speaking, is still a single-particle
transport. There the charge current vanishes for zero voltage
while the spin current remains, leading in this way to the
appearance of a persistent spin current across the junction.

For the Cooper-pair tunneling the charge and spin Josephson currents
are determined by $I_J^{\rm charge} = 2e ~{\rm Im} [e^{-2eVt/\hbar}
\Phi^{\rm charge}(eV)] $ and $I_J^{\rm spin} = 2e~ {\rm Im}
[e^{-2eVt/\hbar} \Phi^{\rm spin}(eV)]$ where
%
\begin{eqnarray}
\lefteqn{\Phi_{\rm charge,spin}(i\omega)=-\frac{1}{\beta}\sum_{\omega_n}\sum_{{\bf
k},{\bf p}} \left[T_{\bf k+q/2, p+q/2} T_{\bf -k+q/2, -p+q/2}
F_1^{\uparrow, \downarrow}({\bf k}, i\omega_n){F_1^{\downarrow,
\uparrow}}^*({\bf p}, i\omega_n-i\omega_m) \right. }
&& \nonumber\\
&& \left. \pm T_{\bf k-q/2, p-q/2} T_{\bf -k-q/2, -p-q/2}
F_1^{\downarrow, \uparrow}({\bf k}, i\omega_n){F_1^{\uparrow,
\downarrow}}^*({\bf p}, i\omega_n-i\omega_m) + T_{\bf k+q/2,
p+q/2}T_{\bf -k-q/2,
-p-q/2} \times \right. \nonumber\\
&& \left. F_2^{\uparrow, \uparrow}({\bf k},
i\omega_n){F_2^{\uparrow, \uparrow}}^*({\bf p}, i\omega_n-i\omega_m)
\pm  T_{\bf k-q/2, p-q/2}T_{\bf -k+q/2, -p+q/2}F_2^{\downarrow,
\downarrow}({\bf k}, i\omega_n{)F_2^{\downarrow, \downarrow}}^*({\bf
p},
i\omega_n-i\omega_m) \right].
\end{eqnarray}
Once more $+$ or $-$ refer to the charge and spin Josephson
current, respectively. Evaluating the sum over Matsubara's
frequencies the charge current can be found
\begin{equation}
\label{joscharge00} I_J^{\rm
charge}(eV)=[J_1(eV)+J_2(eV)\cos\Delta\varphi]\sin(\Delta\theta+2eVt)
+[J_3(eV)+J_4(eV)\cos\Delta\varphi]\cos(\Delta\theta+2eVt) \quad,
\end{equation}
where the explicit expressions of the coefficients $J_1(eV)$, and
$J_2(eV)$ are given as
\begin{eqnarray}
\lefteqn{J_1(eV)  =  2 e|T|^2 \sum_{{\bf k},{\bf p},i \neq j}
\frac{|\Delta_{\bf q}|^2}{\sqrt{\varepsilon^2_{{\bf
k}s}+|\Delta_{\bf q}|^2}
\sqrt{\varepsilon^2_{{\bf p}s}+|\Delta_{\bf q}|^2}} \times } && \nonumber \\
&& \left[\left( 1 - \frac{\varepsilon_{{\bf k}a}\varepsilon_{{\bf
p}a}}{\sqrt{\varepsilon^2_{{\bf k}a}+|h_{\bf
q}|^2}\sqrt{\varepsilon^2_{{\bf p}a}+|h_{\bf q}|^2}}\right) \left(
\frac{1}{eV+E_{i{\bf k}}+E_{i{\bf p}}}- \frac{1}{eV-
E_{i{\bf k}}-E_{i{\bf p}}} \right)\right. +  \nonumber \\
&& \left. \left( 1 + \frac{\varepsilon_{{\bf k}a}\varepsilon_{{\bf
p}a}}{\sqrt{\varepsilon^2_{{\bf k}a}+|h_{\bf
q}|^2}\sqrt{\varepsilon^2_{{\bf p}a}+|h_{\bf q}|^2}}\right) \left(
\frac{1}{eV+E_{i{\bf k}}+E_{j{\bf p}}}- \frac{1}{eV -E_{i{\bf
k}}-E_{j{\bf p}}} \right)\right] \quad, \label{joscharge0}
\end{eqnarray}
and
\begin{eqnarray}
\lefteqn{J_2(eV)  =  2e|T|^2 \sum_{{\bf k},{\bf p},i \neq j}
\frac{|\Delta_{\bf q}|^2}{\sqrt{\varepsilon^2_{{\bf
k}s}+|\Delta_{\bf q}|^2} \sqrt{\varepsilon^2_{{\bf
p}s}+|\Delta_{\bf q}|^2}}  \left(\frac{|h_{{\bf
q}}|^2}{\sqrt{\varepsilon^2_{{\bf k}a}+|h_{\bf
q}|^2}\sqrt{\varepsilon^2_{{\bf p}a}+|h_{\bf q}|^2}}\right) \times } && \nonumber \\
&& \left[  \left( \frac{1}{eV+E_{i{\bf k}}+E_{j{\bf p}}}-
\frac{1}{eV- E_{i{\bf k}}-E_{j{\bf p}}} \right) - \left(
\frac{1}{eV+E_{i{\bf k}}+E_{i{\bf p}}}- \frac{1}{eV-E_{i{\bf
k}}-E_{i{\bf p}}} \right)\right] \quad. \label{21}
\end{eqnarray}
$J_3(eV)$ and $J_4(eV)$ are found similarly,
\begin{eqnarray}
\lefteqn{J_3(eV)  =  2 \pi e|T|^2 \sum_{{\bf k},{\bf p},i \neq j}
\frac{|\Delta_{\bf q}|^2}{\sqrt{\varepsilon^2_{{\bf
k}s}+|\Delta_{\bf q}|^2}
\sqrt{\varepsilon^2_{{\bf p}s}+|\Delta_{\bf q}|^2}} \times } && \nonumber \\
&& \left\{\left( 1 - \frac{\varepsilon_{{\bf k}a}\varepsilon_{{\bf
p}a}}{\sqrt{\varepsilon^2_{{\bf k}a}+|h_{\bf
q}|^2}\sqrt{\varepsilon^2_{{\bf p}a}+|h_{\bf q}|^2}}\right) \left[
\delta(eV-E_{i{\bf k}}-E_{i{\bf p}})- \delta(eV+
E_{i{\bf k}}+E_{i{\bf p}}) \right]\right. +  \nonumber \\
&& \left. \left( 1 + \frac{\varepsilon_{{\bf k}a}\varepsilon_{{\bf
p}a}}{\sqrt{\varepsilon^2_{{\bf k}a}+|h_{\bf
q}|^2}\sqrt{\varepsilon^2_{{\bf p}a}+|h_{\bf q}|^2}}\right) \left[
\delta(eV-E_{i{\bf k}}-E_{j{\bf p}})- \delta(eV +E_{i{\bf
k}}+E_{j{\bf p}}) \right]\right\} \quad, \label{12}
\end{eqnarray}
and
\begin{eqnarray}
\lefteqn{J_4(eV)  =  2 \pi e|T|^2 \sum_{{\bf k},{\bf p},i \neq j}
\frac{|\Delta_{\bf q}|^2}{\sqrt{\varepsilon^2_{{\bf
k}s}+|\Delta_{\bf q}|^2} \sqrt{\varepsilon^2_{{\bf
p}s}+|\Delta_{\bf q}|^2}}  \left(\frac{|h_{{\bf
q}}|^2}{\sqrt{\varepsilon^2_{{\bf k}a}+|h_{\bf
q}|^2}\sqrt{\varepsilon^2_{{\bf p}a}+|h_{\bf q}|^2}}\right) \times } && \nonumber \\
&& \left[  \delta(eV-E_{i{\bf k}}-E_{j{\bf p}})- \delta(eV+
E_{i{\bf k}}+E_{j{\bf p}}) -  \delta(eV-E_{i{\bf k}}-E_{i{\bf p}})
+ \delta(eV+E_{i{\bf k}}+E_{i{\bf p}}) \right] \quad.
\label{joscharge}
\end{eqnarray}
%
Since $J_3(0)=0$ and $J_4(0)=0$, Eq. (\ref{joscharge}) becomes for $V=0$
\begin{equation}
\label{joscharge-v=0}
I^{\rm charge}_J(0)=[J_1(0)+J_2(0)\cos\Delta\varphi]\sin\Delta\theta.
\end{equation}
The above result is identical, up to the precise
expressions of the current amplitudes, to the zero voltage result
of Ref. \onlinecite{Kulic}, though there the superconducting order
parameter is uniform. Thus, a nonzero voltage in helimagnetic
FFLO-like superconductors affect the Josephson current in an
essential way.

For the Josephson spin current we find:
%
\begin{equation}
\label{spinc0} I_J^{\rm spin}(eV)= \sin\Delta\varphi \left[ \tilde
J_1(eV) \cos(\Delta\theta +2eVt)+\tilde
J_2(eV)\sin(\Delta\theta+2eVt)  \right]\quad.
\end{equation}
where

\begin{eqnarray}
\lefteqn{\tilde J_1(eV)   =    2 \mu_B|T|^2 \sum_{{\bf k},{\bf p}}
\frac{|\Delta_{\bf q}|^2|h_{\bf q}|^2 }{\sqrt{\varepsilon^2_{{\bf
k}s}+|\Delta_{\bf q}|^2} \sqrt{\varepsilon^2_{{\bf
p}s}+|\Delta_{\bf q}|^2} \sqrt{\varepsilon^2_{{\bf k}a}+h_{\bf
q}|^2}\sqrt{\varepsilon^2_{{\bf p}a}+h_{\bf q}|^2}} \times} &&
\nonumber \\
&&  \left[ \frac{1}{eV +E_{i{\bf k}}+E_{i{\bf p}}}- \frac{1}{eV
  -E_{i{\bf k}}-E_{i{\bf p}}} -    \left( \frac{1}{eV
+E_{i{\bf k}}+E_{j{\bf p}}}- \frac{1}{eV - E_{i{\bf k}} -E_{j{\bf
p}}} \right) \right]
 \label{spinc}
\end{eqnarray}
and
\begin{eqnarray}
\lefteqn{\tilde J_2(eV)   =   2 \pi \mu_B|T|^2 \sum_{{\bf k},{\bf
p}} \frac{|\Delta_{\bf q}|^2|h_{\bf q}|^2
}{\sqrt{\varepsilon^2_{{\bf k}s}+|\Delta_{\bf q}|^2}
\sqrt{\varepsilon^2_{{\bf p}s}+|\Delta_{\bf q}|^2}
\sqrt{\varepsilon^2_{{\bf k}a}+h_{\bf
q}|^2}\sqrt{\varepsilon^2_{{\bf p}a}+h_{\bf q}|^2}} \times} &&
\nonumber \\
&&  \left[ \delta(eV -E_{i{\bf k}}-E_{i{\bf p}})- \delta(eV
 +E_{i{\bf k}}+E_{i{\bf p}}) -     \delta(eV
-E_{i{\bf k}}-E_{j{\bf p}}) + \delta(eV + E_{i{\bf k}}+E_{j{\bf
p}}) \right]
 \label{spinc1}
\end{eqnarray}
\end{widetext}
 At zero voltage the spin Josephson current becomes
\begin{equation}
\label{spinc-0v}
I_J^{\rm spin}(0)   =\tilde J_1(0)  \cos\Delta \theta\sin \Delta \varphi.
\end{equation}
We see that  the term proportional to $\sin
\Delta \varphi \cos\Delta \theta$ vanishes for zero voltage because
$\tilde J_2(0)=0$, similarly to the charge Josephson current case.
This result also agrees with the corresponding one in Ref. \onlinecite{Kulic}.
Note once more the crucial role played by the voltage in this system.

\section{Effect of an external magnetic field}

Interesting results follow from Eqs. (\ref{joscharge-v=0}) and (\ref{spinc-0v}) in the
presence of an external magnetic field ${\bf H}$ perpendicular to the current
direction, say $x$-direction, and in the plane of the junction. By
assuming that the external field points in the $z$-direction, it is
straightforward to derive the results
\begin{equation}
I_J^{\rm charge}(0)=[J_1(0)+J_2(0)\cos\Delta\varphi]\sin\left(\Delta\theta
+\frac{2\pi Hyl}{\Phi_0}\right),
\end{equation}
\begin{equation}
I_J^{\rm spin}(0)   = J(0) \cos\left(\Delta\theta
+\frac{2\pi Hyl}{\Phi_0}\right)\sin \Delta \varphi,
\end{equation}
where $l=2\lambda+d$, with $\lambda$ being the penetration depth and
$d$ the junction thickness. $\Phi_0$ is the elementary flux quantum.
Indeed, the magnetic field can only couple to the phase of the
superconducting order parameter. The helimagnetic order parameter
is {\it neutral} and for this reason its phase cannot couple to
the external magnetic field. If the junction has a cross-section of
area $L_yL_z$, the total currents
$I^{\rm charge}_{\rm J,tot}$ and
$I^{\rm spin}_{\rm J,tot}$ flowing through the junction is
obtained by integrating the $y$-variable over the interval $[0,L_y]$:
\begin{equation}
\label{I-H-c}
I^{\rm charge}_{\rm J,tot}=
(I_{1c}+I_{2c}\cos\Delta\varphi)\frac{\Phi_0}{\pi\Phi}
\sin\left(\frac{\pi\Phi}{\Phi_0}\right)
\sin\left(\Delta\theta+\frac{\pi\Phi}{\Phi_0}\right),
\end{equation}
\begin{equation}
\label{I-H-s}
I^{\rm spin}_{\rm J,tot}=I_c
\frac{\Phi_0}{\pi\Phi}\sin\left(\frac{\pi\Phi}{\Phi_0}\right)
\cos\left(\Delta\theta+\frac{\pi\Phi}{\Phi_0}\right)\sin\Delta\varphi,
\end{equation}
where $I_{1c}=J_1(0)L_yL_z$, $I_{2c}=J_2(0)L_yL_z$,
$I_c=J(0)L_yL_z$, and $\Phi=HL_yl$.

From Eqs. (\ref{I-H-c}) and (\ref{I-H-s}) we see that the
phase difference $\Delta\theta$ can be adjusted in such a way to
vanish the spin Josephson current (\ref{I-H-s}). Remarkably, also
the opposite situation is
possible, i.e., the vanishing of the charge Josephson current by
adjusting the phase difference $\Delta\theta$. This constitutes an
example of a system with a spin current but no charge current.

\section{Conclusion}

We have shown that a dissipationless dc and ac
Josephson spin currents exist between two non-uniform
superconductors with helimagnetic order. For the spin current the nonzero average
$\langle c_{\uparrow}^{\dagger} c_{\downarrow}\rangle$ plays a crucial role.
We expect that effects similar to the ones discussed here may also happen in some
Superconductor/Ferromagnet/Insulator/Ferromagnet/Superconductor
(SFIFS) heterostructures.\cite{Kulic,Kriv,BuzdinRMP}

Measuring a spin current is presently a considerable challenge.
One way could be to detect the electric fields induced by such a
current.\cite{flavio,loss} However, in our case the signature of
the spin current would be a corresponding modulation of the charge
Josephson current as follows from Eq.(\ref{joscharge}).

One of the main results of our analysis was discussed in Sect. IV, namely,
the possibility of using an external magnetic field to tune the charge
Josephson current to zero, while the spin Josephson current remains
non-vanishing. In such a situation the resulting spin Josephson effect is
very similar to the one discussed recently in the context of
Ferromagnet/Ferromagnet tunnel junctions.\cite{flavio} However, here
we have a much better control of the system through the external magnetic
field. This result presumably holds also in the case of a helimagnetic
superconductor with a uniform order parameter.\cite{Kulic}

The model we have studied here assumes the
coexistence of a FFLO state with helimagnetism.
In order to fully confirm the validity of this scenario,
further theoretical and experimental investigation is necessary.

\acknowledgments

We are thankful to P. Fulde and M. Titov for the helpful
discussions. We would like to thank in
particular M. L. Kuli\'c for his valuable comments.
I.E. and F.S.N. would like to thank the
Laboratoire de Physique des Solides, Universit\'e Paris-Sud, where
part of this work has been done, for the hospitality.

\end{document}